# Performance Analysis and Dynamic Evolution of Deep Convolutional Neural Network for Nonlinear Inverse Scattering

Lianlin Li, *Senior Member*; Long Gang Wang, and Fernando L. Teixeira, *Fellow*

*Abstract*—The solution of nonlinear electromagnetic (EM) inverse scattering problems is typically hindered by several challenges such as ill-posedness, strong nonlinearity, and high computational costs. Recently, deep learning has been demonstrated to be a promising tool in addressing these challenges. In particular, it is possible to establish a connection between a deep convolutional neural network (CNN) and iterative solution methods of nonlinear EM inverse scattering. This has led to the development of an efficient CNN-based solution to nonlinear EM inverse problems, termed DeepNIS. It has been shown that DeepNIS can outperform conventional nonlinear inverse scattering methods in terms of both image quality and computational time. In this work, we quantitatively evaluate the performance of DeepNIS as a function of the number of layers using structure similarity measure (SSIM) and mean-square error (MSE) metrics. In addition, we probe the dynamic evolution behavior of DeepNIS by examining its near-isometry property. It is shown that after a proper training stage the proposed CNN is near optimal in terms of the stability and generalization ability.

## I. Introduction

The solution of nonlinear electromagnetic (EM) inverse scattering problems is of interest in a number of applications [1-10]. These solutions are able to take into account multiple scattering effects inside the probed scene [1]. Many inverse scattering algorithms have been developed over the years [4-5]; however, their application to large, realistic scenes is hampered by high computational costs. In recent years, deep convolutional neural networks (CNN) have shown to be a promising tool for solving inverse problems due to the increasing availability of very large data sets and the concomitant increase in the available computational power [11,12]. For example, CNN-based strategies have been successfully applied in magnetic resonance imaging, X-ray computed tomography [13], and computational optical imaging [14,15]. It has been found that they can typically outperform conventional image reconstruction techniques in terms of improved image quality and computational speed [16].

Manuscript received MM DD, YY; revised MM DD,YY; accepted MM DD, YY. This research was funded in part by National Natural Science Foundation of China (NSF) grant 61471006.
L. Li and L. Wang are with the School of Electronics Engineering and Computer Sciences, Peking University, Beijing, 100871, China
lianlin.li@pku.edu.cn
F. L. Teixeira is with the ElectroScience Laboratory, The Ohio State University, Columbus OH, 43212, USA; teixeira.5@osu.edu.

More recently, Li et al. [17] investigated the connection between deep CNNs and iterative methods for nonlinear EM inverse scattering. Based this connection, they proposed a complex-valued CNN architecture for tackling nonlinear EM inverse scattering, termed `DeepNIS'. A complex-valued CNN is a straightforward extension of conventional real-valued CNN [11,12]. DeepNIS is a non-iterative solver, which greatly reduces the computational costs compared to iterative techniques. In parallel to this, Wei and Cheng [18] have recently applied a U-net-based deep neural network to nonlinear EM inverse scattering problems, where three different input methods have been comprehensively studied.

In this work, we quantitatively evaluate the performance of DeepNIS as a function of the number of layers using the structure similarity measure (SSIM) and mean-square error (MSE) metrics. The performance is evaluated using the MNIST dataset [19]. In addition, we probe the dynamic evolution behavior of DeepNIS by examining its near-isometry property. After a proper training stage this probing study shows that the proposed CNN is near optimal in terms of the stability and generalization ability. The results show the potential of DeepNIS in tackling nonlinear inverse scattering problems.

## II. Theoretical Foundations

This section summarizes the connection between the CNN architecture and Born iterative method for solving nonlinear EM inverse scattering problems. Throughout this work, the time dependence factor $\exp(-i\omega t)$ is implied and omitted.

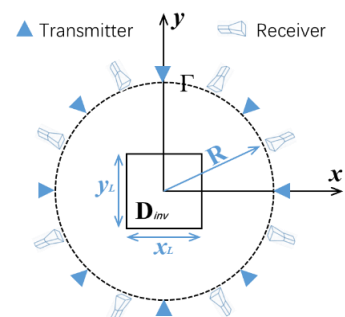

**Fig.1.** Data acquisition setup for the EM inverse problem.

We illustrate our strategy in a 2D multiple-input multiple-output data acquisition setup in Fig. 1. The investigation domain $\Omega$ is successively illuminated by TM-polarized incident waves $E_{in}(r;\omega,\theta)$ (with $\theta$ being the index of the $\theta$ th illumination). The transmitters and receivers are both located in the observation domain $\Gamma$. For the $n$th illumination ($n$=1, 2, …, $N$) and the $m$th ($m$=1, 2, …, $M$) receiver, the scattered electrical field $E_{sca}^{(n)}$ at $r_m$ is obtained by the coupled equations [5-8]:

$$E_s(r;\omega,\theta) = \int_\Omega G(r,r';\omega)E_\theta(r';\omega)\chi(r')dr' \quad (1)$$

$$E_\theta(r;\omega,\theta) = E_{in}(r;\omega,\theta) + \int_\Omega G(r,r';\omega)E_\theta(r';\omega)\chi(r')dr', \quad (2)$$

where $r' = (x',y')$ and $r = (x,y)$ denote source and observation points, respectively, and $G(r,r')$ is the free-space Green's function. In addition, $\chi = k^2/k_0^2 - 1$ is the contrast function, where $k$ and $k_0$ are the wavenumbers of the probed object and background medium, respectively.

For numerical implementation, the investigation domain $\Omega$ is discretized into a series of pixels and the discrete field values represented by column vectors. After that, Eqs. (1) and (2) can be rewritten in compact form as:

$$E_{sca}^{(n)} = G_d E^{(n)} \chi \quad (3)$$

and

$$E^{(n)} - E_{inc}^{(n)} = G_s E^{(n)} \chi \quad (4)$$

*Algorithm I.* Born iterative solution.

---
$n = 0; E_\theta^{(n)} = E_\theta^{(inc)}$
WHILE NOT convergence
● $\chi^{(n+1)} = argmin_\chi \sum_\theta \|E_{s,\theta} - G_d diag(E_\theta^{(n)})\chi\|_2^2$    (5)
● Updating $E_\theta^{(n+1)}$ from the state equation (4).
$n \leftarrow n + 1$
END WHILE

---

The Born iterative method can be applied to solve the nonlinear EM inverse problem described by Eqs. (3) and (4) as summarized in *Algorithm I*. The critical stage is to solve the time-consuming and ill-posed optimization problem represented by Eq. (5). We attempt to address this difficulty by exploring the CNN strategy. To that end, we modify Eq. (5) as follows:

$$\chi^{(n+1)} = argmin_\chi \left\{ \sum_\theta \xi_\theta^{(n)} \|E_{s,\theta} - G_d diag(E_\theta^{(n)})\chi\|_2^2 + \mathcal{R}(\chi) \right\} \quad (6)$$

where $\{\xi_\theta^{(n)}\}$ denote illumination-dependent weighting factors used to adjust the contributions from different measurements. The regularization term $\mathcal{R}(\chi)$ incorporates any prior on the contrast function $\chi$ and is used to mitigate the ill-posedness of the inverse problem. By applying so-called one-step first-order gradient-based approach [3] to solve Eq. (6), we obtain *Algorithm II*, where $A_\theta^{(n)} = G_d diag(E_\theta^{(n)})$. We assume that $E_\theta^{(n)}$ can be statistically approximated by its nearest stationary field $\hat{E}_\theta^{(n)}$. This implies that $(\hat{E}_\theta^{(n)})^* \hat{E}_\theta^{(n)}$ is shift-invariant and thus $w_\theta^{(n)} = I + \eta_\theta (A_\theta^{(n)})^* A_\theta^{(n)}$ behaves like a typical convolutional kernel. The iterative index $n$ can be understood as the layer index of the deep neural network, while the soft-threshold function $\mathcal{S}\{\cdot\}$ corresponds to the nonlinear activation function in deep learning [17]. In this sense, we establish the connection between the Born-iterative method and deep CNN, where $\hat{p}^{(n)}(\theta)$ extracts the illumination-dependent features. By comparing this strategy to conventional iterative inverse scattering methods, the expectation is that the learning method would be more efficient as it optimizes the weighting matrices and biases, and targets the reconstruction error with respect to the ground-truth images. The above observations suggest that deep CNN networks are naturally well-suited for nonlinear EM inverse scattering problems.

*Algorithm II*. Modified Born iterative solution.

---
$n = 0, p^{(0)} = \chi^{(n)}$
FOR $n = 1,2,…,K$
● $\hat{p}^{(n)}(\theta) = \mathcal{S}\left\{ \left(I + \eta_\theta (A_\theta^{(n)})^* A_\theta^{(n)}\right)\chi^{(n)} - \eta_\theta (A_\theta^{(n)})^* E_{s,\theta} \right\}$
● $\chi^{(n)} = \sum_\theta \xi_\theta \hat{p}^{(n)}(\theta)$
● Updating $E_\theta^{(n+1)}$.
END FOR

---

### III. EXPERIMENTAL RESULTS

In the following, we evaluate DeepNIS performance for solving nonlinear EM inverse scattering problems. To evaluate reconstructed image quality, we adopt the structure similarity index metric (SSIM) and mean-square error (MSE) metrics. In addition, we examine its dynamic evolution.

*III.A Simulation setup*

We train and test the DeepNIS using the MNIST dataset [17,19]. With reference to Fig. 1, the region of interest $D_{inv}$ is a square domain of size $5.6\lambda_0 \times 5.6\lambda_0$ ($\lambda_0$=7.5 cm is the working wavelength in vacuum), which is uniformly divided into $110 \times 110$ pixels for the simulations. The MNIST dataset elements shown in the first column of Figure 2a constitute randomly placed targets within each sample. A total of 36 linearly polarized transmitters uniformly spaced over the circumference $\Gamma$ with radius R=$10\lambda_0$ are used to successively illuminate the investigation domain. At the same time, 36 co-polarized receivers are used to collect the scattered electric field. The MNIST dataset elements are assumed to be lossless dielectrics with relative permittivity $\varepsilon_r = 3$. In addition, 30 dB white noise is added to the results of the full-wave forward simulations [20]. A total of $10^4$ image pairs constituted by back-propagated images and original (ground truth) images randomly chosen from the MNIST dataset are divided into three sets: 7000 image pairs for training, 1000 for validation, and 2000 for blind testing. The training stage is done using the ADAM optimization method [21], with mini-batches size of 32, and epoch setting as 50. The learning rates are set to $10^{-4}$ and $10^{-5}$ for the first two layers and the last layer in each network, respectively, and halved once the error plateaus. The complex-valued weights and biases are initialized by random weights with zero mean Gaussian distribution and standard deviation of $10^{-3}$. The computations are performed with AMD Ryzen Threadripper 1950X 16-Core processor, NVIDIA GeForce GTX 1080Ti, and 128GB access memory. The networks are designed using the Tensor Flow library [22].

*III.B Numerical results*

Figure 2(a) shows the images obtained by using the proposed CNN with the numbers of layers varying from one to nine. For reference, the corresponding ground truths and the Born-iterative method (BIM) results are provided in the first column and second column, respectively. Throughout this work, the total BIM iterative number is set to be 25, because after which no further visible improvement on the reconstruction quality can be observed. These results clearly illustrate that well-trained CNNs with four layers or more can produce excellent reconstructions. In contrast, the Born iterative method provides relatively poorer reconstructions in this case. Figs. 2(b) and (c) examine quantitatively the quality of the images by using the proposed CNNs with different number layers in terms of the SSIM and MSE metrics. Fig. 2(b) plots the dependence of the averaged SSIMs and MSEs as a function of the number of CNN layers. The averaged SSIMs and MSEs are computed over 2000 training samples and 2000 test samples. In Fig. 2(b), a sample of image reconstructions of the digit "5" is provided for reference. Fig. 2(c) reports the statistical histograms of the image quality in terms of SSIM corresponding to CNN with one, three, five, seven, and nine layers, respectively, over 2000 test images. The *y*-axis is normalized with respect to the total 2000 test images. Based on these results, several conclusions can be made. First, when the CNN has more than five layers, both MSEs and SSIMs converge to a stable level. It can be clearly seen that the DeepNIS with more than five layers can match the ground truth results very well. Second, if the CNN has more layers, the resultant discrepancy between training and testing performance increases. This is likely a consequence of the fact that deeper CNNs require more network parameters to be optimized over a given training sample and hence more prone to over-fitting. Finally, it is worth mentioning that it only takes a well-trained DeepNIS less than one second to construct an image in this case, whereas it takes Born iterative algorithm about nearly one hour. Based on the above results, it can be concluded that a properly trained DeepNIS solution clearly outperforms the Born iterative method in terms of both image quality and computation time.

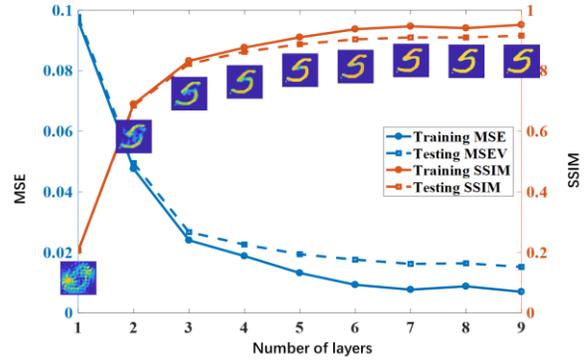

(b)

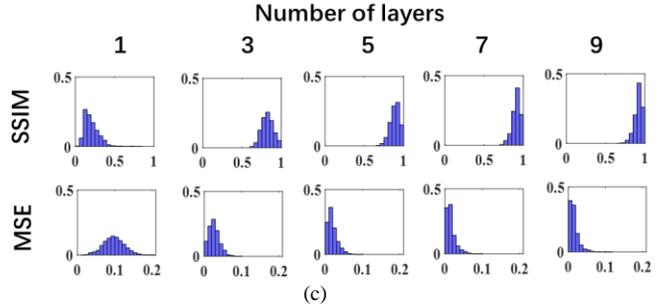

(c)

**Fig.2**. Digit-like object reconstructions with relative permittivity $\varepsilon_r = 3$ by different EM inverse scattering methods. (a) Reconstructed images obtained by using the proposed CNN with varying the numbers of layers from one to nine. The corresponding ground truth data and the reconstructed image obtained by using the Born-iterative method are shown in the first column and second column, respectively. (b) Dependence of the averaged SSIMs and MSEs as function of the number of CNN layers. (c) Statistical histograms of the image quality in terms of SSIM and MSE. Here, 2000 test samples are used.

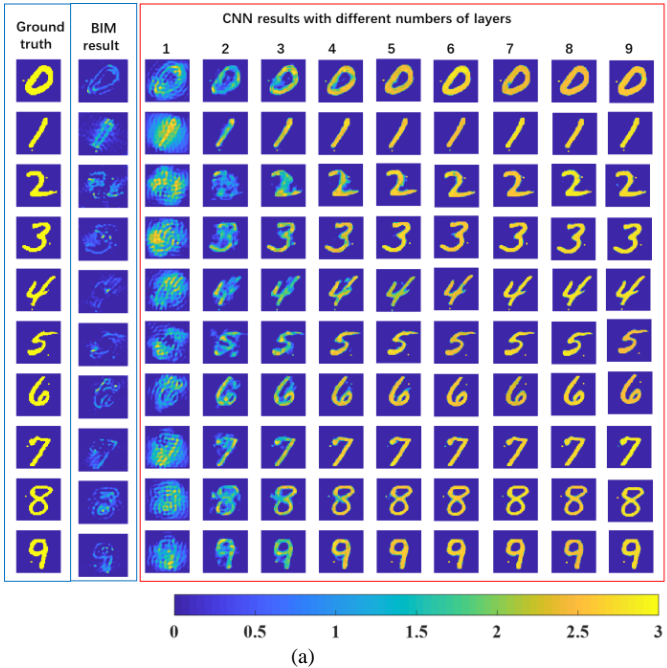

Figure 3 provides further insights into the CNN-based EM inverse scattering solution. In this case, the number of CNN layers is fixed at five and the ground truth corresponds to a digit "2"-like object. In this set of figures, nine randomly selected features extracted at different CNN depths are illustrated. It can be seen that with an increase in CNN depth, the extracted features approach gradually the ground truth. It can be conjectured that the extracted features more or less reflect the contrast function $\hat{p}^{(n)}(\theta)$ under the different illumination conditions. In other words, these results suggest the CNN-based approach is not merely matching patterns but actually has a learning capability to represent the underlying nonlinear inverse electromagnetic scattering problem.

*III.C Dynamic Evolution Behavior of DeepNIS*

Stability and generalization are two crucial issues for deep neural networks. Here, we examine the evolution behavior of feature propagation through the proposed CNN. More formally, the feature at the output layer of the CNN can be described as $\chi^{(out)} = \mathcal{F}(\chi^{(in)}; \Theta)$, where $\Theta$ encapsulates the parameters of the network block. Inspired by the restricted isometry property widely examined in compressive sensing [3], we focus on the parameter $\eta = \|\delta\chi^{(out)}\|_2^2 / \|\delta\chi^{(in)}\|_2^2$, where $\delta\chi^{(in)}$ and $\delta\chi^{(out)}$ denotes the input perturbation to the CNN network and the associated output perturbation, respectively. Evidently, the network with $\eta \gg 1$ is unstable since the output feature would be highly sensitivity to data perturbations. Conversely, if $\eta \ll 1$ the corresponding network block lacks generalization capability since input features would be exponentially suppressed and not properly identified at the output. Consequently, an "optimal" neural network should have $\eta \approx 1$. In other words, a deep neural network with $\eta \approx 1$ is near optimal in terms of stability and generalization. Fig. 4 plots the dependence of averaged $\eta$ as the function of iteration number for the CNN with different number layers (and with the other parameters set as before). It can be seen that the proposed CNNs with more than three layers, which are trained after ten iterations, indeed have $\eta \approx 1$.

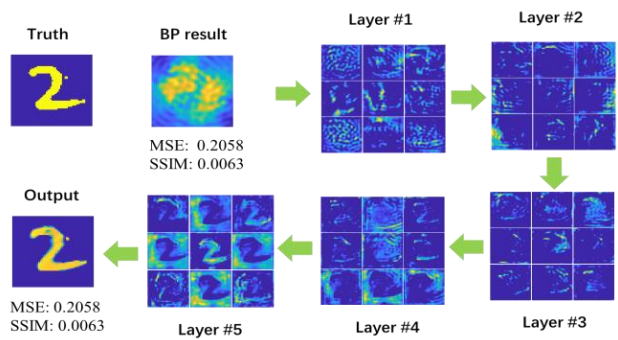

**Fig.3**. Nine randomly selected features extracted at different CNN depths.

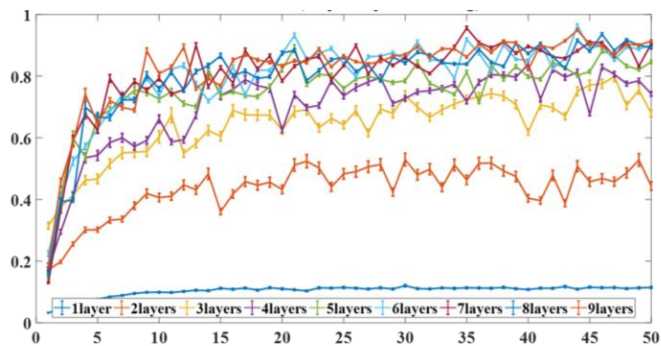

**Fig.4**. Dependence of averaged $\eta$ with respect to the iterative epoch number for the CNNs with different number of layers.

## IV. CONCLUSIONS

We have evaluated the performance of DeepNIS quantitatively as a function of the number of layers based on different quantitative metrics. We have shown that DeepNIS shows advantages over conventional inverse scattering methods in terms of image quality and computational time. We have also investigated the dynamic evolution of DeepNIS. The analysis shows that following a proper training stage the proposed CNN is near optimal in terms of the stability and generalization ability. Together, these results indicate the clear potential of DeepNIS in tackling nonlinear inverse scattering problems. It is plausible that more advanced or tailored CNN architectures could yield better results, which will be explored in a future study.